# The Effects of Major League Baseball's Ban on Infield Shifts: A Quasi-Experimental Analysis

Lee Kennedy-Shaffer, PhD

*Department of Biostatistics, Yale School of Public Health, New Haven, CT, USA*

Correspondence: Lee.kennedy-shaffer@yale.edu

## Abstract

From 2020 to 2023, Major League Baseball changed rules affecting team composition, player positioning, and game time. Understanding the effects of these rules is crucial for leagues, teams, players, and other relevant parties to assess their impact and to advocate either for further changes or undoing previous ones. Panel data and quasi-experimental methods provide useful tools for causal inference in these settings. I demonstrate this potential by analyzing the effect of the 2023 shift ban at both the league-wide and player-specific levels. Using difference-in-differences analysis, I show that the policy increased batting average on balls in play and on-base percentage for left-handed batters by a modest amount (nine points). For individual players, synthetic control analyses identify several players whose offensive performance (on-base percentage, on-base plus slugging percentage, and weighted on-base average) improved substantially (over 70 points in several cases) because of the rule change, and other players with previously high shift rates for whom it had little effect. This article both estimates the impact of this specific rule change and demonstrates how these methods for causal inference are potentially valuable for sports analytics—at the player, team, and league levels—more broadly.

**Introduction**

Professional sports provide seemingly endless opportunities for data-driven adaptation, as teams and players identify and exploit potential advantages. The leagues that govern these sports are not immune to these pressures and have changed rules, formats, and broadcast schedules in attempts to increase viewership, relevance, and profits. For example, the National Football League has instituted multiple changes to kick-off and tackling rules to try to balance player safety and excitement (Pennington 2018; Sullivan 2024) and the National Basketball Association has changed player rest rules (Marks 2023). Major League Baseball (MLB) recently implemented several changes around team makeup, pitcher hitting, pitch timing, baserunning, and fielding play, largely designed to reduce game times and/or increase offense and the number of baserunners (Doolittle 2022; Kennedy-Shaffer 2022; Mains 2024).

Analyzing the effects of these rules—on both their intended and unintended outcomes—should be approached with the same degree of statistical rigor as the analysis of players and teams. However, for a variety of reasons, this has generally not been the case. There is less incentive for this analysis as there is limited benefit to anyone other than the league; even if it does occur by either leagues or teams, it is unlikely to be public. Public analyses using formal causal inference methods are rare (Markes et al. 2024). In addition, there is a statistical challenge: top-flight leagues are one-of-a-kind and therefore the core requirement of repetition and multiple observations is difficult.

Panel data methods and so-called quasi-experimental causal inference can provide a solution to the statistical problem, given that certain key assumptions are met. These frameworks, which are often used in policy and economic evaluation, comprise a variety of methods with different assumptions and data requirements (see, e.g., Basu et al. 2017; Craig et al. 2017; Cunningham 2021; Huntington-Klein 2022). They generally rely on longitudinal or panel data sets, where the outcome is measured repeatedly, in some cases before and in some cases after the policy goes into effect. The simplest approach, a pre-post design or interrupted time series, compares the post-policy outcome to the pre-policy outcome, either using the time points closest in time to each other, averages before and after the change, or a model of the time trend (Craig et al. 2017). This approach, although not often stated as such, is commonly used in public-facing articles around rules changes; see, e.g. Mains (2024). The major downsides are that it assumes a consistent and predictable time trend in the outcome and cannot disentangle the policy of interest from any other changes occurring at the same time.

To address these drawbacks, controlled pre-post designs can be used if there are time series available both for observations that are affected by the policy change and for observations that are not affected (Craig et al. 2017). One such approach, difference-in-differences (DID), requires the affected and unaffected observations or groups to have parallel time trends in the counterfactual world where no policy change occurs, although the time trend need not be explicitly modeled (Basu et al. 2017). Another approach, the synthetic control method (SCM), requires multiple unaffected time series; none of these needs to parallel the time trend of the affected time series. Rather, the method identifies a weighted average of those unaffected groups that is a good match for the affected group (Basu et al. 2017). These methods have been used in settings as diverse as identifying the effects of minimum wage increases (Card and Krueger 1994), geopolitical case studies across countries (Abadie et al. 2015), and, recently, the impact of COVID-19 interventions (Cowger et al. 2022) and vaccines (Kennedy-Shaffer 2024), among many others. More detailed discussions of the requirements and implementations of these methods are available in textbooks and academic articles such as Cunningham (2021), Huntington-Klein (2022), Caniglia and Murray (2020), Abadie et al. (2010), and Abadie (2021).

In sports, these methods have become popular in evaluation of the economic and social impacts of sports events, teams, or stadiums (Bradbury 2022; García Bulle et al. 2022; Kobierecki and Pierzgalski 2022; Pyun 2019; Suggs et al. 2024) and in business analyses of revenue and attendance effects of decisions (Brook 2022; Cardazzi and Rodriguez 2024; Cisyk 2020). A recent study also looked at the effect of MLB's automatic baserunner rule in extra innings on game times using DID (Kennedy-Shaffer 2022). Another study used both DID and SCM to analyze the effect of the 1981 change in match result point values on competitive balance in the English Premier League (Sharma 2024). As far as I am aware, however, these methods have not previously been used on game play outcomes using team, league, and player statistics (other causal methods have; see, e.g., Markes et al. (2024)). This provides an opportunity to use modern statistical methods to fully understand the impact of these rule changes.

In this paper, I study the effect on batting outcomes of the recent MLB rule banning the infield shift (Mains 2024). Specifically, the rule mandates that the defensive team have two players on each side of second base in the infield. I look at both league-level and player-level outcomes, which provide estimates of effects of interest to different audiences: namely the league and team business operations, who are most interested in the rule's ability to increase "exciting" game events by getting more baserunners; and team baseball operations and the players and their representatives, who may be interested in the rule's effect on their own game and how to change their playing style, team-building approach, or signing decisions in response. I also discuss the broader potential for the use of these methods in sports analytics, as well as open questions that will inform for which applications, settings, and outcomes they can most beneficially be used.

**Materials and Methods**

This section summarizes the methods used. More details are available in Appendix B, and all data and code are included in the GitHub repository ([https://bit.ly/QE-Baseball](https://bit.ly/QE-Baseball)).

*Analysis 1: League-Wide Effects (Difference-in-Differences)*

To analyze league-wide effects of the rule change, we are interested in comparing shift situation plate appearances (PAs) to non-shift situation PAs. Previous cross-sectional (within a single season) analyses have compared shifted and non-shifted PAs or—for a less biased comparison—shifted vs. non-shifted PAs among players with substantial numbers of each (Carleton 2018a; b; Petriello 2018). Of course, after the shift ban, there are no shifted PAs, so we cannot use shifted and non-shifted PAs as our time series. Other analyses have compared right-handed to left-handed batters, given that left-handed batters are shifted much more frequently (Mains 2024; Pavitt 2024), performing analyses similar to DID when looking at the effect of the ban. Here, I use this approach with a full recent time series, and limit the analysis to bases-empty PAs, since shifts are more common in those situations. Data are compiled from the FanGraphs "Splits Leaderboard" (2024) across MLB, by batter handedness for PAs with no runners on.

Using these data, I conduct a DID analysis comparing left-handed batters (LHBs) to right-handed batters (RHBs) in these PAs, after vs. before the shift ban (i.e., in the 2023 season vs. the 2022 season). The same analysis is conducted for earlier seasons (2015–2022, skipping the shortened 2020 season) compared to one (full) season prior, where we should see no effect because no policy change occurred. It is also conducted for the 2024 season compared to 2023, which would indicate the difference in the effect of the rule in its second season compared to the first. The 2015 season was chosen as the starting point for two reasons: first, it is the beginning of full Statcast data on fielder positioning, which will be relevant for other analyses; second, it is a reasonable starting point for the modern shift era where all teams were employing it to some degree (Carleton 2018a). For this analysis, where the emphasis is on league-wide effects and MLB may be the most relevant audience, I consider two outcomes: batting average on balls in play (BABIP) and on-base percentage (OBP), which relate to MLB's goals for the policy change of increasing base hits and having more baserunners (Arthur 2022; Carleton 2022a).

### Analysis 2: Player-Specific Effects (Synthetic Control Method)

To analyze the effects of the policy change on specific players, we turn to player-specific data and the synthetic control method. First, season-average or -total batting statistics were collected for all players with at least 250 PAs in any of the 2015–2024 seasons, excluding the shortened 2020 season, via Baseball Savant's "Statcast Custom Leaderboard" (2024). Shift rates for all players who had at least 250 PAs in each of 2021–2023 were obtained from Baseball Savant's "Statcast Batter Positioning Leaderboard" (2024). Names were matched using the R package baseballr (Petti and Gilani 2024). These players were categorized into low (no more than 15% of PAs), high (at least 75%) and medium (15–75%) shift rates based on their 2022 shift rate in bases-empty PAs. This resulted in 30 players in the high-shift category; these target players will be the focus of the analysis. The 58 players in the low-shift category form the controls or "donor pool" for the SCM analysis. Here, I consider three batting outcomes that are useful in assessing the batting value of the player and would be relevant directly to him, his team, and his representatives: OBP, on-base plus slugging (OPS), and weighted on-base average (wOBA).

For each of the 30 target players, I add to the 2021–2023 data the seasons from 2015–2019 in which they had at least 250 PAs. Their donor pool is obtained by finding the players in the overall donor pool who had at least 250 PAs in those seasons. For each outcome, a synthetic control is fit using the time series data prior to 2023. The outcome, player's age, PAs, hits, singles, home runs, base-on-balls percentage, and strikeout percentage are used as the covariates. All covariates except the outcome itself are used individually for 2021 and 2022 and averaged over all included pre-2020 seasons; the outcome is used individually for each season while age is only included as a 2022 covariate. This yields a weighted average of players in the donor pool (i.e., low-shift players) whose history of the specified outcome (i.e., OBP, OPS, or wOBA) is most similar to the target player. Specifically, the method identifies weights (restricted to be between 0 and 1 and sum to 1) for the control players that minimize the squared distance from a weighted average of the covariates, where the covariate weights are chosen to best predict the outcome trajectory in the pre-intervention period. See Appendix B and Abadie (2021) for further technical details of the synthetic control method; it is implemented using the tidysynth R package (Dunford 2023).

The synthetic control weights are then applied to the donor players' 2023 results to get a synthetic (or counterfactual) outcome for the target player: what we would expect their 2023 outcome to be if the rule had not been changed. This is subtracted from their actual 2023 outcome to get an estimate of the effect of the rule on their outcome.

I then perform placebo tests by obtaining null distributions for each outcome against which to compare the effect estimates. The same SCM analysis is conducted for each of the 58 low-shift control players as if they were the target player, excluding them from the donor pool. This gives a placebo or null distribution of the estimates we would see among players if there were no effect of the rule change, capturing the natural uncertainty of the weighting algorithm and the year-to-year fluctuations in batting performance.

To evaluate the persistence of effects, the same analysis is conducted, limited to players with at least 250 PAs in both 2023 and 2024. This yields a subset of the above player list (27 target players and 42 control players). The synthetic control is re-fit for these groups and estimated effects obtained for both 2023 and 2024.

Additional robustness checks are conducted by performing two similar analyses where zero or near-zero effects are expected. The first—an "in-unit" placebo analysis—conducts the same analysis for 2023 effects, using as the target players the 25 players who otherwise fit the criteria for inclusion with a shift rate in 2022 between 15 and 30%. We would expect these players to see little to no effect of the rule change when compared to the true placebo players. The second conducts a similar analysis as the original using 2022 as a dummy intervention year for the target players. This "in-time" placebo analysis can provide some sense of any bias of using these two groups of players as the targets and controls. It would, however, be expected to have more variability as there is less data that it can use to fit the synthetic control.

## Results

### *Analysis 1: League-Wide Effects (Difference-in-Differences)*

The results of the league-wide DID analysis for BABIP and OBP with bases empty are given in Table 1. Both indicate a modest increase in the outcome for left-handed batters in 2023 compared to what would have been expected in the absence of a rule change, with both estimates (coincidentally) equal to 0.009, or 9 points. To situate this magnitude in terms of yearly fluctuations and assess the parallel trends assumption, Figure 1 plots the trends in the two outcomes by batter handedness and year, beginning in 2015 and excluding 2020 (A,B) and the same analysis for each of these seasons compared to the previous full season (C,D). Other batting summary outcomes (average, slugging percentage, OPS, wOBA) had similar results (see the interactive R Shiny app at https://bit.ly/SCM-baseball for all results). The estimated effect of the rule change for both base-on-balls and strikeout percentage was an increase, with a higher magnitude for walks (0.5 percentage points) than strikeouts (0.1).

*Table 1.* Two-by-two DID analysis for the effect of the shift ban in 2023, comparing left-handed and right-handed batters' BABIP and OBP with the bases empty for 2023 vs. 2022. Data source: FanGraphs "Splits Leaderboard" (2024).

| Batter Handedness | BABIP | | | OBP | | |
|---|---|---|---|---|---|---|
| | **2022** | **2023** | *Difference* | **2022** | **2023** | *Difference* |
| **LHB** | 0.275 | 0.287 | *0.012* | 0.299 | 0.315 | *0.015* |
| **RHB** | 0.291 | 0.294 | *0.003* | 0.303 | 0.309 | *0.006* |
| *Difference* | *−0.016* | *−0.007* | *0.009* | *−0.004* | *0.006* | *0.009* |

For years where no policy change occurred, the DID estimate centered around 0, generally with a small magnitude. This indicates the larger 2023 estimate likely represents a true effect. However, both 2021 and 2022 had estimates (pre-trends) less than 0. This may reflect a divergence in the trends of RHBs and LHBs, possibly due to increasing shifts and more effective fielder positioning or other rule changes with a differential effect (like the universal designated hitter). If anything, this may indicate that LHBs would have continued to worsen compared to RHBs, causing the 2023 estimates to underestimate the positive effect of the change.

*Figure 1.* Trends by batter handedness (A,B) and DID estimates (C,D) for BABIP (A,C) and OBP (B,D) by year, 2015–2023, excluding 2020, for bases empty plate appearances. The counterfactual 2023 value is also shown (A,B) assuming the increase from 2022 would be the same for LHBs as for RHBs in the absence of the rule change. Each DID estimate (C,D) is a comparison of LHBs to RHBs for the listed season compared to the previous full season. Data source: FanGraphs "Splits Leaderboard" (2024). Results for other outcomes are available at https://bit.ly/SCM-baseball.

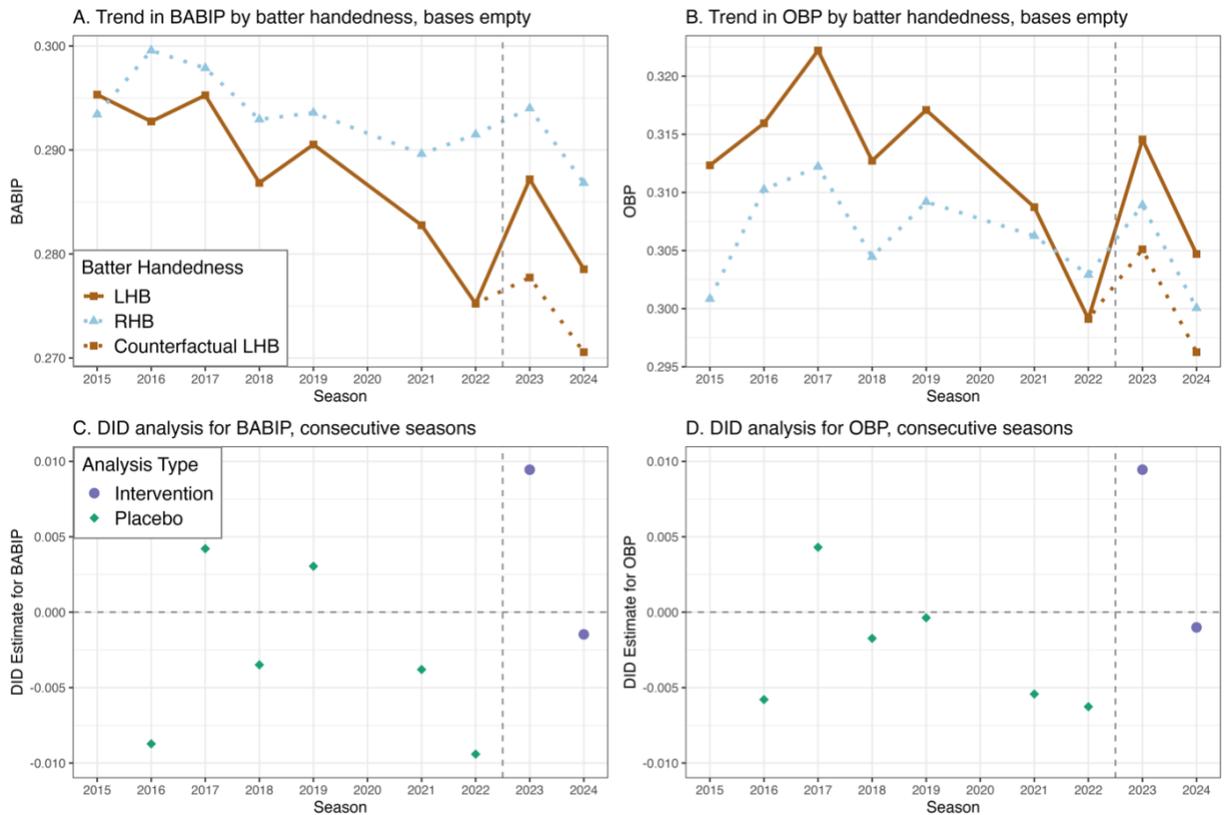

In 2024, effect estimates compared to 2023 returned to nearly 0 (–0.001 to be precise). This indicates that the 2023 effect persisted in 2024, and further suggests the 2023 effect was due to the policy change rather than other time trends causing divergence between left- and right-handed batters. The exceptions were walk and strikeout percentage, which both had modest negative effects in 2024, suggestive of a possible change in pitching approach for LHBs in 2024 compared to 2023.

***Analysis 2: Player-Specific Effects (Synthetic Control Method)***

To illustrate the player-specific SCM estimates, I show results here for Corey Seager. Seager had over 650 PAs and the highest shift rate (92.8%) of any of the target players in 2022. He was also specifically noted as a player likely to benefit from the new rule (Petriello 2023b). That analysis relied on modelling his batted-ball locations in 2022 and identifying balls in play that would have likely been hits in the absence of the shift. While valuable, especially in predicting effects *a priori*, that analysis cannot account for changes in approach (by both pitcher and batter) in the shift-ban environment or for other changes in the batting environment between 2022 and 2023. The SCM accounts for these changes but relies on the identified weighted average being a good comparison for the target player, so there is value in conducting both analyses and triangulating results as a causal inference strategy (Matthay et al. 2020).

The players with non-negligible weights in the SCM fits for Corey Seager for all three outcomes are shown in Table 2, along with their weights. The relative consistency of included players, sparsity of the number of players with non-zero weights, and explicit weight values provide transparency of the method. This allows the investigator to assess the validity of the weights using other qualitative or quantitative information. In this case, for example, Trea Turner has a high similarity score with Corey Seager ("Corey Seager" 2024), which may provide plausibility to the results.

*Table 2.* Top-weighted donor players in the synthetic control fits for Corey Seager, for the three outcomes. Players with weights below 0.1% are excluded.

| Weight Ranking | OBP | | OPS | | wOBA | |
|---|---|---|---|---|---|---|
| 1 | Starling Marte | 63% | Trea Turner | 60% | Trea Turner | 47% |
| 2 | Carlos Correa | 37% | Carlos Correa | 20% | Jean Segura | 25% |
| 3 | - | | Jean Segura | 10% | Carlos Correa | 21% |
| 4 | - | | Yan Gomes | 6% | José Abreu | 8% |
| 5 | - | | Paul Goldschmidt | 5% | - | |

Seager's synthetic OBP in 2023 was estimated to be 0.305, compared to an observed OBP of 0.390, for an estimated effect of the rule change of 85 points. For OPS, the synthetic value was 0.742 and the observed 1.013, for an estimated effect of 271 points. For wOBA, the synthetic value was 0.304 and the observed 0.419, for an estimated effect of 115 points. In all cases, the estimates point to a positive effect of the rule change on Seager's performance, increasing his offensive output by more than 25% compared to what it would have been under the old rules. The trajectories of observed and synthetic control outcomes for Seager are shown in Figure 2.

The results for all target players are given in Table 3; individual player results can be explored in the interactive R Shiny app (https://bit.ly/SCM-baseball). The estimated effects of the rule change for each player for each of the three outcomes are included, as well as a p-value based on the placebo tests. This indicates the proportion of the placebo test estimates (and that target estimate) that the target estimate tied or exceeded in absolute value. Because of the relatively small number of donor players for some target players, the coarseness of this type of placebo test, and the large numbers of results, these are more useful as an indicator of the reliability of the result rather than as strict hypothesis tests. Plots of the estimates for each of the three outcomes for all target and donor pool players, as well as their pre-intervention fit, are given in Figure 3.

Overall, for each of the three outcomes, most target players (over 75%) had an estimated increase. The placebo test estimates, as expected, had estimates greater than 0 near 50% of the time (although it was somewhat higher for OPS and wOBA, indicating possible systemic underestimation of the 2023 outcomes). The mean and median value of the estimates across the target players were also much larger (4–8 times) than those for the placebo test players.

**Figure 2.** Synthetic and observed outcomes (OBP, OPS, wOBA) for Corey Seager compared to the synthetic control constructed by donor players with shift rates no higher than 15% in 2022. Constructed using full (non-2020) seasons from 2016–2023 excluding seasons where Seager had fewer than 250 PAs (i.e., 2018). Donor players had at least 250 PAs in the same seasons. Data sources: Baseball Savant's "Statcast Custom Leaderboard" (2024) and "Statcast Batter Positioning Leaderboard" (2024). Results for other players are available at https://bit.ly/SCM-baseball.

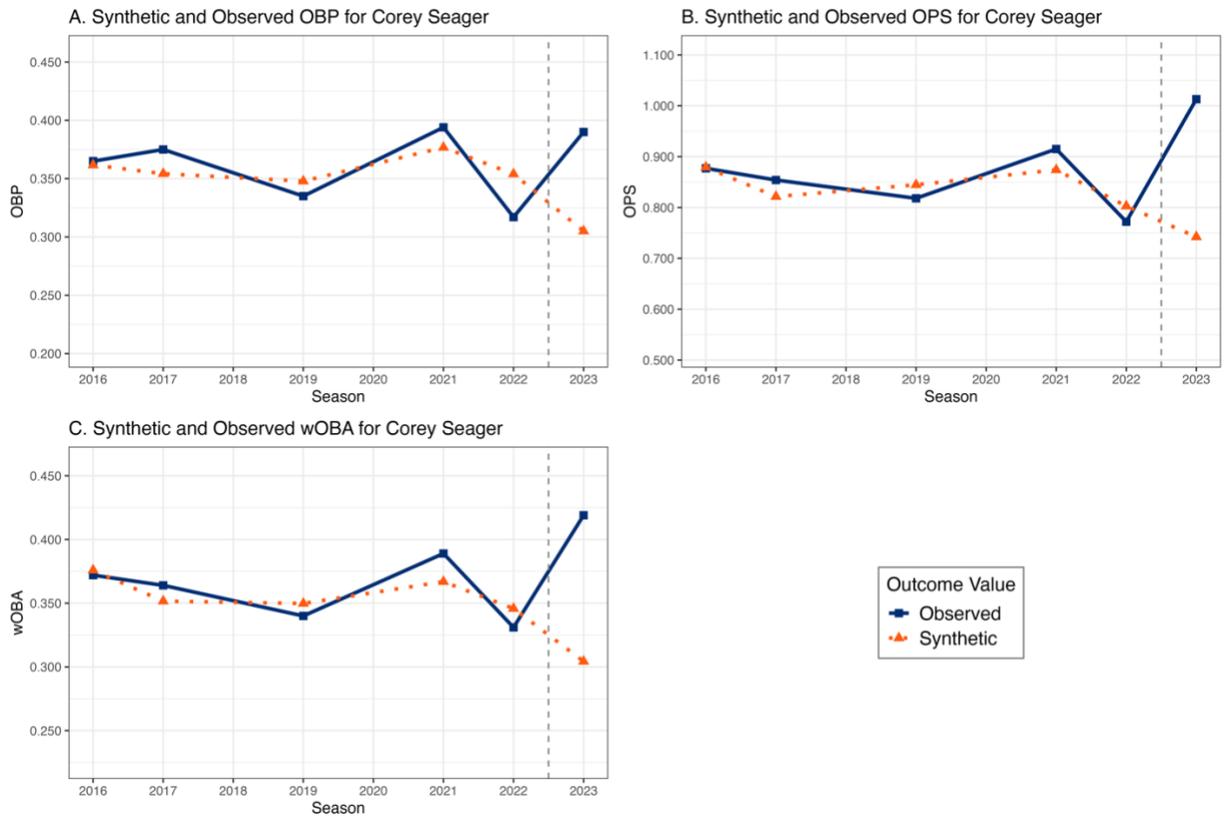

*Table 3.* Shift rates in 2022 and estimated effects of the shift ban in 2023 on OBP, OPS, and wOBA, with associated placebo test p-values, for players with at least an 75% shift rate in 2022 and at least 250 PAs in each of the 2021–2023 seasons. Data sources: Baseball Savant's "Statcast Custom Leaderboard" (2024) and "Statcast Batter Positioning Leaderboard" (2024).

| Player | Shift Rate (2022) | OBP | | OPS | | wOBA | |
|---|---|---|---|---|---|---|---|
| | | Est. | P | Est. | P | Est. | P |
| Corey Seager | 92.8% | 0.085 | 0.017 | 0.271 | 0.017 | 0.115 | 0.017 |
| Kyle Tucker | 90.9% | 0.047 | 0.136 | 0.097 | 0.254 | 0.034 | 0.254 |
| Kyle Schwarber | 90.7% | 0.018 | 0.424 | 0.061 | 0.424 | 0.026 | 0.39 |
| Cody Bellinger | 90.5% | 0.041 | 0.169 | 0.13 | 0.102 | 0.047 | 0.119 |
| Joey Gallo | 90.0% | −0.001 | 0.932 | 0.037 | 0.644 | 0.018 | 0.593 |
| Max Kepler | 89.7% | 0.023 | 0.356 | 0.096 | 0.254 | 0.044 | 0.119 |
| Max Muncy | 89.0% | 0.005 | 0.831 | −0.026 | 0.695 | −0.005 | 0.881 |
| Seth Brown | 88.6% | −0.035 | 0.22 | −0.043 | 0.627 | −0.021 | 0.525 |
| Shohei Ohtani | 88.3% | 0.076 | 0.017 | 0.241 | 0.017 | 0.083 | 0.017 |
| Yordan Alvarez | 88.1% | 0.015 | 0.458 | 0.214 | 0.017 | 0.083 | 0.017 |
| Brandon Lowe | 85.3% | 0.012 | 0.627 | 0.078 | 0.373 | 0.033 | 0.271 |
| Brandon Belt | 85.2% | 0.044 | 0.136 | 0.115 | 0.169 | 0.047 | 0.119 |
| Eddie Rosario | 83.6% | 0.001 | 0.966 | 0.001 | 1 | 0.003 | 0.949 |
| Cavan Biggio | 82.8% | 0.015 | 0.475 | 0.03 | 0.678 | 0.013 | 0.644 |
| Anthony Rizzo | 82.6% | −0.005 | 0.797 | −0.05 | 0.559 | −0.013 | 0.712 |
| Matt Olson | 81.3% | 0.075 | 0.017 | 0.222 | 0.017 | 0.085 | 0.017 |
| Mike Yastrzemski | 81.2% | 0.009 | 0.678 | 0.078 | 0.373 | 0.034 | 0.254 |
| Eugenio Suárez | 80.9% | 0.002 | 0.915 | −0.077 | 0.373 | −0.013 | 0.678 |
| Byron Buxton | 78.8% | −0.022 | 0.356 | 0.074 | 0.373 | 0.02 | 0.559 |
| Rowdy Tellez | 78.4% | −0.024 | 0.339 | −0.083 | 0.339 | −0.017 | 0.61 |
| Carlos Santana | 78.2% | −0.004 | 0.864 | 0.021 | 0.78 | 0.005 | 0.881 |
| Jorge Soler | 78.0% | 0.021 | 0.356 | 0.081 | 0.339 | 0.028 | 0.356 |
| José Ramírez | 77.3% | 0.058 | 0.051 | 0.109 | 0.203 | 0.046 | 0.119 |
| Josh Naylor | 77.1% | 0.031 | 0.271 | 0.136 | 0.085 | 0.047 | 0.119 |
| Joc Pederson | 77.0% | 0.024 | 0.339 | −0.006 | 1 | 0.001 | 0.983 |
| Daniel Vogelbach | 76.4% | 0 | 0.983 | −0.013 | 0.898 | −0.04 | 0.169 |
| Bryce Harper | 75.7% | 0.043 | 0.153 | 0.094 | 0.254 | 0.04 | 0.169 |
| Salvador Perez | 75.6% | −0.02 | 0.407 | 0.049 | 0.559 | 0.017 | 0.593 |
| Marcus Semien | 75.3% | 0.022 | 0.356 | 0.158 | 0.085 | 0.037 | 0.22 |
| Ozzie Albies | 75.1% | 0.03 | 0.271 | 0.165 | 0.068 | 0.063 | 0.068 |

*Figure 3.* Plots of estimated effects of the shift ban rule change in 2023 with pre-intervention trends in the difference between synthetic control estimated and observed statistic for OBP (A), OPS (B), and wOBA (C). All included players had at least 250 PAs in each of the 2021–2023 seasons; target players had at least a 75% shift rate in 2022 while placebo players had no more than a 15% shift rate in 2022. Data sources: Baseball Savant's "Statcast Custom Leaderboard" (2024) and "Statcast Batter Positioning Leaderboard" (2024).

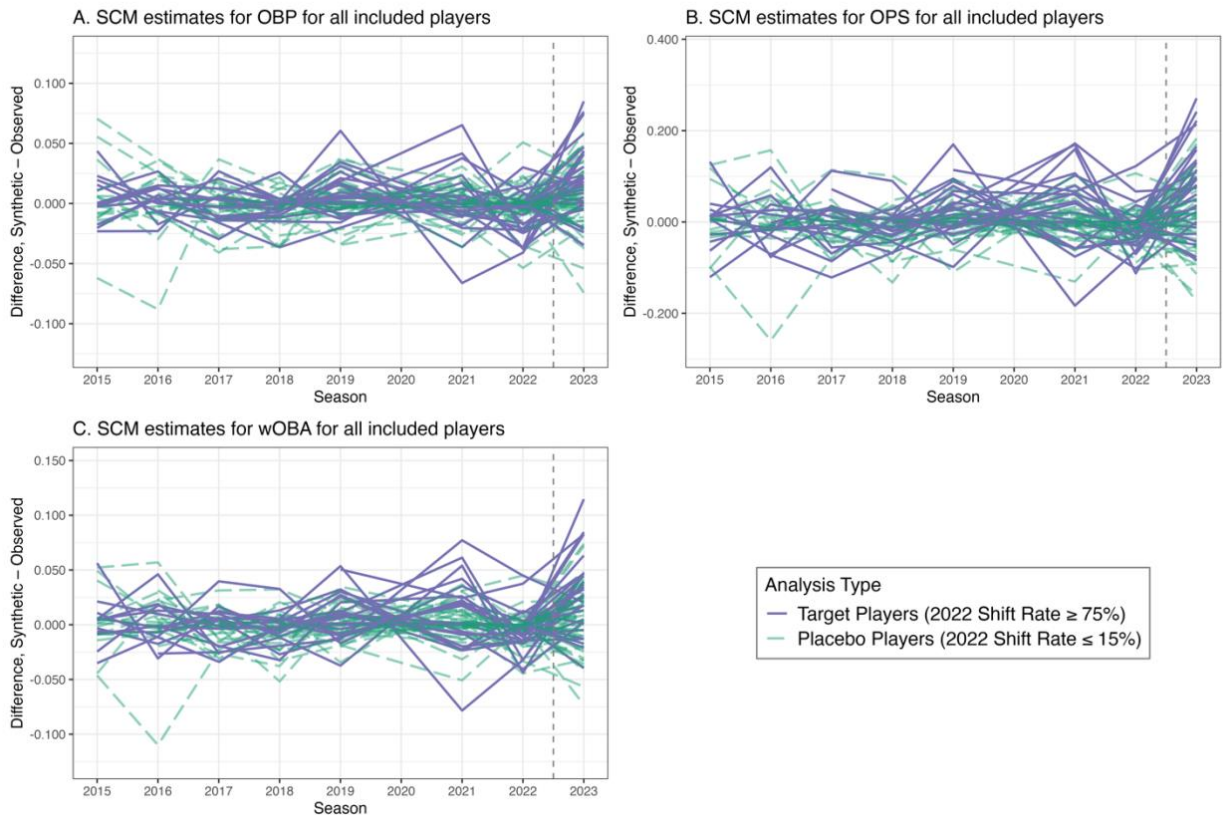

Figure 4 plots the effect estimates for the target players versus the player's 2022 shift rate, indicating some dose response in the effect of the shift ban. Namely, players shifted more frequently in 2022 exhibited higher estimated effects of the ban in 2023. This relationship was modest, however, and the players' responses exhibit high variability. On average, a 10 percentage point increase in 2022 shift rate corresponds to an increase in estimated effect for the player by 11 points in OBP, 31 points in OPS, and 17 points in wOBA.

**Figure 4.** Plots of estimated effects of the shift ban rule change in 2023 for target players vs. the player's 2022 shift rate for OBP (A), OPS (B), and wOBA (C). All included players had at least 250 PAs in each of the 2021–2023 seasons; target players had at least a 75% shift rate in 2022 while control players had no more than a 15% shift rate in 2022. Dashed lines indicate ordinary least squares best fit lines. Data sources: Baseball Savant's "Statcast Custom Leaderboard" (2024) and "Statcast Batter Positioning Leaderboard" (2024).

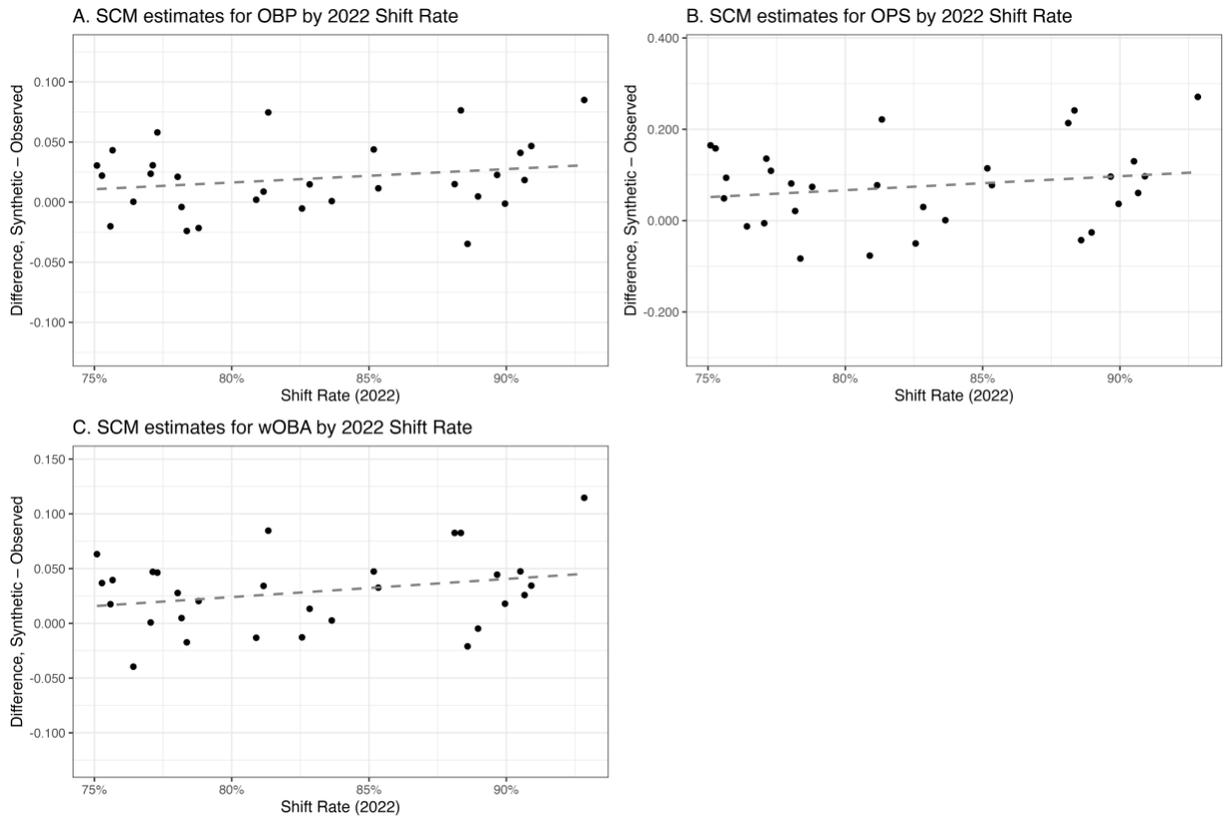

To extend these results to the 2024 season, the SCM results were re-fit, only using the 27 target players and 42 control players who had at least 250 PAs in each of the 2021–2024 seasons. These results are shown in Appendix A, Figures A.1 and A.2, for Corey Seager and all players, respectively. For the total estimated effect in the 2023 and 2024 seasons across the players as whole, the results were similar to the 2023 estimate alone, although larger in magnitude and with more player-to-player variation. This suggests an overall persistence of the effect on the target players.

The in-unit robustness check, conducting the same analysis for the players with 2022 shift rates between 15 and 30%, found a near-zero effect, as expected. The outcomes for these players tended to be very similar to that for the true placebo controls, with neither a consistent positive nor negative effect estimate. The in-time robustness check identified, albeit with high variance, moderate negative effect estimates for many players. This indicates a lack of systematic positive overestimation for these players. It may also reflect the trend, as seen in the pre-intervention DID analyses, of increasing shift usage leading to declining batter performance for left-handed (especially high-shift) batters. Analogues to Figure 3 for these robustness checks are shown in Appendix A, Figures A.3 and A.4, respectively.

**Discussion**

In 2023, MLB instated a ban on infield shifts among a suite of rules changes, with the hope of increasing offense and restoring the look of the game (Arthur 2022; Carleton 2022a). Analyses before and after the implementation of the ban generally relied on one or more of the following approaches: pre-post analyses of overall trends (Mains 2024; Pavitt 2024; Petriello 2023c); cross-sectional comparisons of right- and left-handed batters or different types of batted balls (Mains 2024; Pavitt 2024); models of batted ball outcomes with specific fielder positioning data (Petriello 2023a; b; c); or within-batter comparisons of shift vs. non-shift PAs (Carleton 2018b, 2022b, 2023; Petriello 2018). These can provide valuable estimates of the effect of shifting, specifically its primary effect on how similarly hit balls or similarly used batters are affected by infield alignment. When they include 2023 data, they are estimating the effect of the shift ban in the 2023 run environment, while pre-ban analyses are estimating its effect in a different environment. Nonetheless, the results generally aligned: the shift caused a modest decrease in BABIP for left-handed batters affected by it, which led to lower batting average, OBP, OPS, and wOBA for them as well.

The only formal causal inference analysis of this question known to the author used a different set of causal methods (matching, inverse probability weighting, and instrumental variables) and only pre-ban data. That analysis found a modest run-prevention effect on targeted left-handed batters (Markes et al. 2024). Again, although focused on different outcomes and estimands, the results qualitatively agree with those found here, providing the triangulation of evidence that is useful for causal inference (Matthay et al. 2020).

Comparing modeled outcomes or shifted vs. non-shifted PAs from the pre-2023 era, however, neglects how other rules changes affect play and ignores second- and third-order effects of the shift ban. Specifically, batters and pitchers will change their approaches in response to fielder placement and league rules. Because of this, batters will get different amounts of playing time and face different situations than they would have otherwise, as teams and managers adjust to the new environment (Arthur 2022; Carleton 2022a; b, 2023). These higher-order effects, while presumably smaller in magnitude, are best captured by quasi-experimental designs like the analyses undertaken here. Unlike other pre-post analyses, by using comparison time series that are differentially affected by the shift ban but (presumably) similarly affected by other rules changes, these methods can even attempt to disentangle the effects of different rules changes.

These analyses found a modest benefit to left-handed batters league-wide in terms of BABIP and OBP (9 points for each). Known as the "average treatment effect on the treated" or ATT, these estimate the effect of the shift ban on these outcomes *for left-handed batters in bases-empty plate appearance in the 2023 season*. This is context-specific and may change over time as teams and players adjust their strategies accordingly. These estimates are generally within the ranges predicted from differences between shifted, shaded, and non-shifted PA results by Russell Carleton's within-batter analysis (Carleton 2022b, 2023). These plate appearances represented 23.3% of the total in 2023, so the total effect, assuming no impact of the rule change on other plate appearances, would be expected to be around 2 points, or 1 additional on-base event for every 500 plate appearances. While analysts suggested that still-permitted fielding strategies, such as "strategic shades," and changes in pitching approaches would compensate for some of the effects of the shift ban (Arthur 2022; Carleton 2022a), the effects persisted in 2024, with no meaningful negative estimate for 2024 to indicate reversion of the effect. This indicates either that those strategic adjustments were made quickly enough to nearly entirely take effect within 2023 or they were insufficient to undo the effects of the rule change.

Player-specific analyses found that many high-shift players saw substantial benefits from the shift ban in 2023; four players (Corey Seager, Matt Olson, Yordan Alvarez, and Shohei Ohtani) had estimated wOBA increases over 80 points and OPS increases over 200 points, for example, with Seager, Olson, and Ohtani also seeing estimated OBP increases at 75 or more points. Another four (Josh Naylor, Cody Bellinger, Brandon Belt, and José Ramírez) had estimated OPS increases over 100 points and OBP increases over 30 points. Again, these are ATTs: in this case, estimated effects of the shift ban on these outcomes *for that specific player in the 2023 season*. The most- and least-affected players somewhat align with analyses based on batted ball modeling (Petriello 2023c) and generally correspond to higher or lower 2022 shift rates. The differences in the results may arise from using different outcomes, full- vs. partial-season data, and the batted ball analysis not accounting for the overall hitting environment and changes made by the player to take advantage of the new rule and opposing pitchers and defenses to try to blunt that effect.

The general scale and direction of effects for the target players persisted in 2024, but with increased player-to-player variability. Some players (e.g., Shohei Ohtani, Cody Bellinger, and Marcus Semien for OBP) had similarly large effect estimates in 2024 as in 2023, while others (e.g., Corey Seager and Matt Olson) had large effect estimates in 2023 but estimates close to zero in 2024. Still other players (e.g., Joey Gallo, Seth Brown, and Rowdy Tellez) showed negative effect estimates in both years. Noticeable changes in the effect between years for a player could indicate one, or a combination, of three things: (1) a true change in the effect, due to changes in batting approach, pitching approach, or fielder positioning within the new rules (i.e., shades); (2) variability unrelated to the rule in either that player or a highly-weighted control player; or (3) bias due to the restricted set of control players leading to a worse fit of the synthetic control or the additional year difference leading to further loss of validity of the synthetic control. While these might be able to be disentangled for particular players by analyzing batted ball outcomes or other statistics and models, it is beyond the scope of this analysis. The third possibility there also points to a limitation of these methods, both generally and for sports contexts specifically: long-term time trends are susceptible to more bias due to other differential effects and due to changes in the composition and available controls.

These analyses rely on important untestable assumptions to have causal interpretations: for DID, the assumption of parallel trends in the absence of the policy change is needed; for SCM, the weights must be stable over time for the player's performance in the absence of the policy change. Both analyses also require the assumption of no spillover; that is, that the policy change did not affect the units we consider unaffected. This is clearly not strictly true, as the policy change could somewhat affect all players. Our proxies (RHBs and low-shift "control" players) are presumably less affected, but this can introduce bias; generally it would result in underestimating the magnitude of the effect. The robustness checks provide some evidence of validity of these analyses; the negative in-time placebo effect estimates for 2022 may also indicate some underestimation of the effect of the policy. Moreover, the ATT estimates these methods produce are not necessarily generalizable. They may change over time for the reasons mentioned above, or due to other policy changes or batting environment changes, and they may be different in different leagues.

In addition, the specific choices made in the SCM analysis can affect the results; this includes the choice to drop players entirely for seasons with fewer than 250 PAs, the covariates in the model, and the linear functional form used. Changes to any of these analysis choices, or to the missing data procedures, could change the validity of the SCM fits. For example, incorporating Statcast bat-tracking data such as bat speed instead of or in addition to PA outcomes as covariates in the SCM models could improve the fit and assumptions. More complex DID and SCM methods could also be useful in addressing assumption violations for this or other sports settings (Abadie 2021; Cunningham 2021; Huntington-Klein 2022; Kennedy-Shaffer 2024). Investigating these open questions (parallel trends, spillover, generalizability, time-varying effects, appropriate covariate selection) not only enables better use of these methods but improves understanding of the game and how to affect it more generally.

Overall, quasi-experimental methods can provide a useful framework for causal inference in sports settings. They allow analysts to estimate causal effects in the specific context, and with transparency and interpretability of the causal assumptions. SCM for player performance has the attractive feature of making clear which players are highly weighted in the control, allowing the assessment of validity using other qualitative or quantitative data. Because of this, it may prove useful in assessing effects of injuries, new tactics, or changes in approach of specific players, in a way that can complement both other analytic approaches and more traditional scouting and comparison approaches. At the bigger picture level, these analyses are particularly useful for assessing rules changes and policies that affect players or teams differently within a league, as they can be used in settings with a small number of affected units (even a single unit, as in the case of leagues). In this way, they can provide a substitute or complement to randomized experiments or pilot studies conducted in other leagues—e.g., the minor leagues (Arthur 2022; Lindbergh and Arthur 2021)—which have their own limitations on generalizability. Further research is needed to understand for which settings and outcomes these methods work particularly well, and for which their assumptions may falter. As MLB and other leagues continue to change the rules of the game, the leagues, teams, and players can all benefit from another approach to understanding those effects.

Acknowledgements: The author wishes to thank the participants of the 2024 Carnegie Mellon Sports Analytics Conference and the judges of its reproducible research competition for their helpful feedback and comments.

Declaration of Interests: The author reports there are no competing interests to declare.

**Appendix A: Supplementary Figures**

*Figure A.1.* Synthetic and observed outcomes (OBP, OPS, wOBA) for Corey Seager through 2024 with synthetic control constructed by donor players with shift rates no higher than 15% in 2022. Constructed using full (non-2020) seasons from 2016–2024 excluding seasons where Seager had fewer than 250 PAs (i.e., 2018). Donor players had at least 250 PAs in the same seasons. Data sources: Baseball Savant's "Statcast Custom Leaderboard" and "Statcast Batter Positioning Leaderboard" (2024).

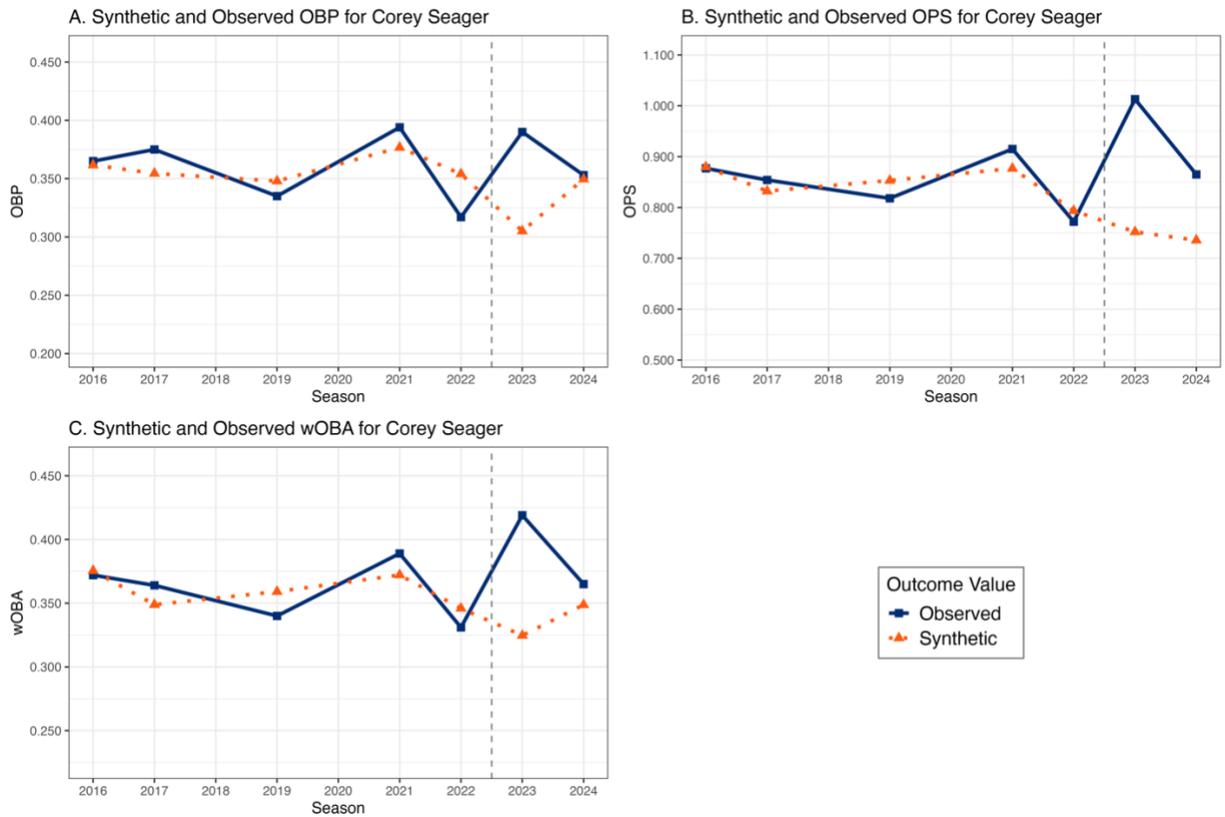

*Figure A.2.* Plots of estimated effects of the shift ban rule change in 2023 and 2024 with pre-intervention trends in the difference between synthetic control estimated and observed statistic for OBP (A), OPS (B), and wOBA (C). All included players had at least 250 PAs in each of the 2021–2024 seasons; target players had at least a 75% shift rate in 2022 while placebo players had no more than a 15% shift rate in 2022. Data sources: Baseball Savant's "Statcast Custom Leaderboard" (2024) and "Statcast Batter Positioning Leaderboard" (2024).

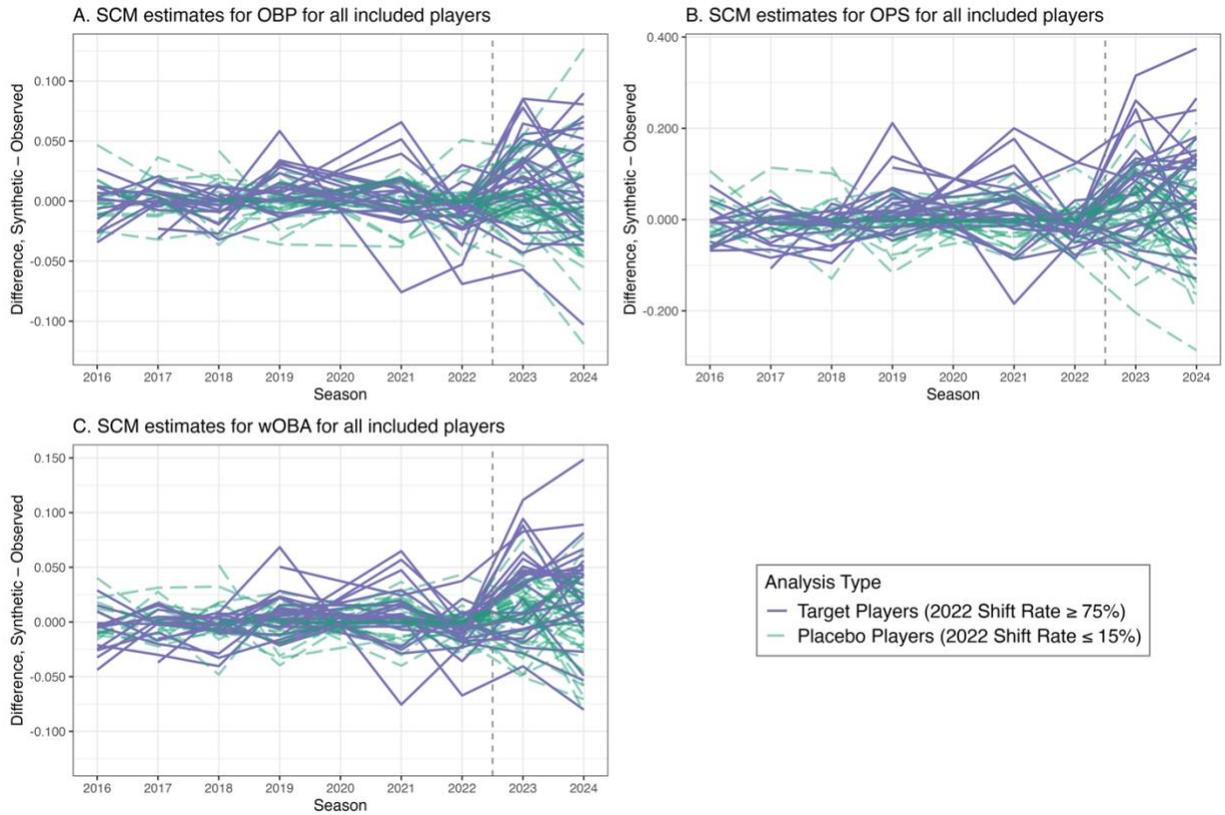

*Figure A.3.* Plots of in-unit placebo estimated effects of the shift ban rule change in 2023 with pre-intervention trends in the difference between synthetic control estimated and observed statistic for OBP (A), OPS (B), and wOBA (C). All included players had at least 250 PAs in each of the 2021–2023 seasons; in-unit placebo players had a 2022 shift rate between 15% and 30% while the true placebo players had no more than a 15% shift rate in 2022. Data sources: Baseball Savant's "Statcast Custom Leaderboard" (2024) and "Statcast Batter Positioning Leaderboard" (2024).

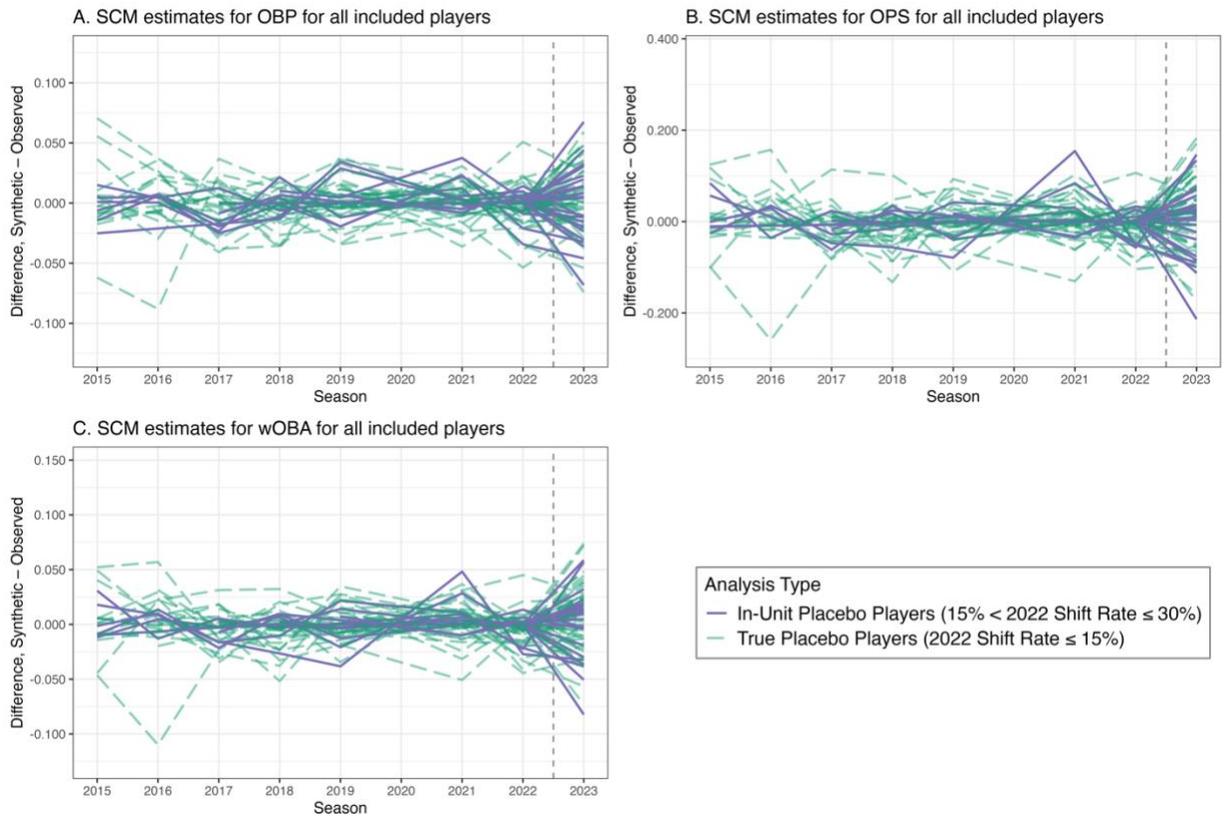

*Figure A.4.* Plots of in-time placebo estimated effects of the shift ban rule change in the pre-intervention year of 2022 with pre-intervention trends in the difference between synthetic control estimated and observed statistic for OBP (A), OPS (B), and wOBA (C). All included players had at least 250 PAs in each of the 2021–2023 seasons; target players had at least a 75% shift rate in 2022 while placebo players had no more than a 15% shift rate in 2022. Data sources: Baseball Savant's "Statcast Custom Leaderboard" (2024) and "Statcast Batter Positioning Leaderboard" (2024).

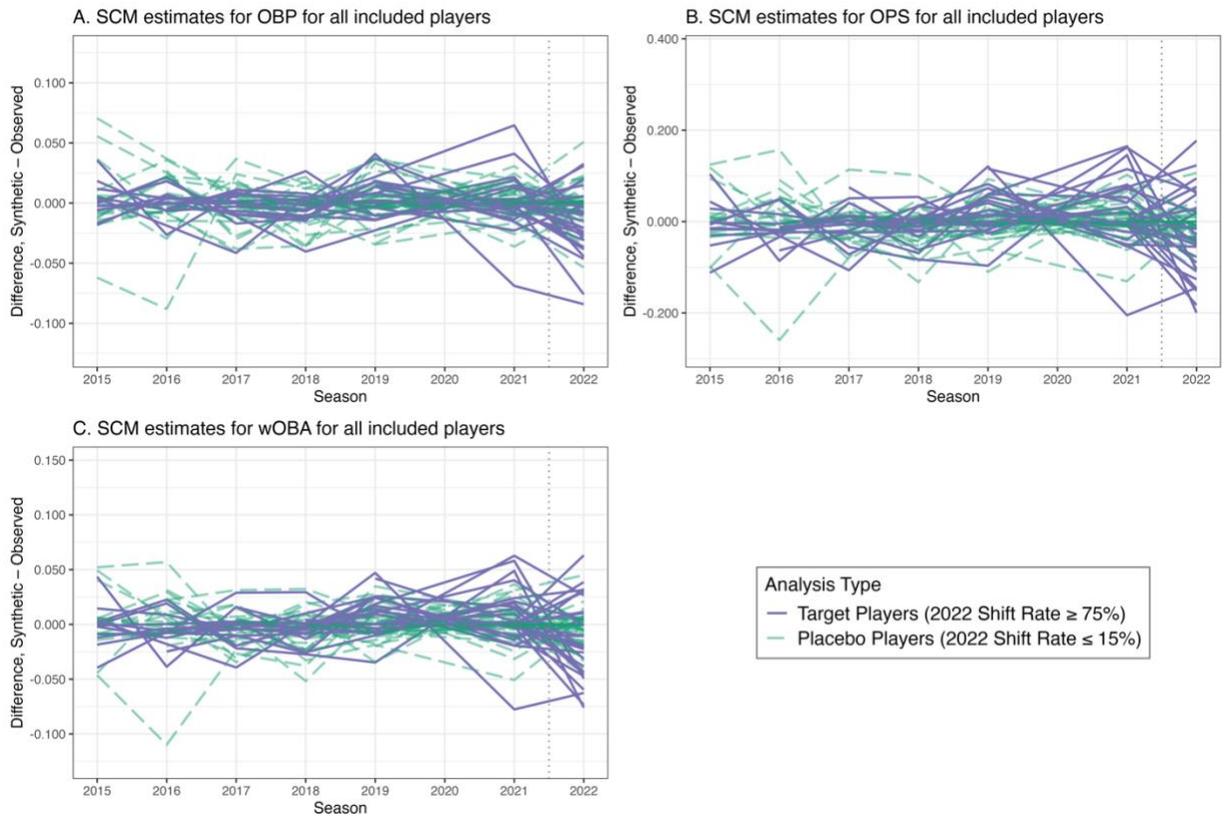

**Appendix B: Technical Appendix**

*Analysis 1: League-Wide Effects (Difference-in-Differences)*

*Notation*

Let $Y_{j,i,t}$ be the value of outcome $j$ (i.e., BABIP or OBP in the main text, batting average, base-on-balls percentage, strikeout percentage, OPS, slugging percentage, or wOBA in the Shiny app) for population $i$ in season $t$, where $i = 1$ corresponds to left-handed batters in bases-empty PAs and $i = 0$ corresponds to right-handed batters in bases-empty PAs, and $t$ ranges from 2015 to 2024, inclusive, except 2020.

Let $Y_{j,i,t}^0$ and $Y_{j,i,t}^1$ be the potential outcomes for outcome $j$, population $i$, season $t$, in the absence and presence, respectively, of the infield shift ban. Let $\theta_{j,i,t} = E[Y_{j,i,t}^1 - Y_{j,i,t}^0]$ be the effect (on the additive scale) of the infield shift ban on outcome $j$ for population $i$ in season $t$. We are interested in the average treatment effect on the treated (ATT) in 2023 and 2024, $\theta_{j,1,2023}$ and $\theta_{j,1,2024}$.

*Assumptions*

(1)  Consistency and no anticipation: For $t < 2023$, $Y_{j,i,t} = Y_{j,i,t}^0$ and for $t \geq 2023$, $Y_{j,i,t} = Y_{j,i,t}^1$ for all $j, i$.

(2)  No spillover: We assume that the infield shift ban has no effect on right-handed batter PAs, so $Y_{j,i,t}^1 = Y_{j,i,t}^0$ for all $j, i, t$.

(3)  Parallel trends: For all $j$, $E[Y_{j,1,t}^0 - Y_{j,1,t-1}^0] = E[Y_{j,0,t}^0 - Y_{j,0,t-1}^0]$ for $t \geq 2023$.

*Estimation*

The estimators are calculated for all $j, t \geq 2016$ by:

$$\hat{\theta}_{j,t} = (Y_{j,1,t} - Y_{j,1,t-1}) - (Y_{j,0,t} - Y_{j,0,t-1}).$$

The implementation can be found at the GitHub repository ([https://bit.ly/QE-Baseball](https://bit.ly/QE-Baseball)).

Under Assumptions (1)–(3) above, $\hat{\theta}_{j,2023}$ is unbiased for $\theta_{j,1,2023}$ for all $j$:

$$E[\hat{\theta}_{j,2023}] = E[(Y_{j,1,2023} - Y_{j,1,2022}) - (Y_{j,0,2023} - Y_{j,0,2022})]$$

$$= E[Y_{j,1,2023}^1 - Y_{j,1,2022}^0] - E[Y_{j,0,2023}^0 - Y_{j,0,2022}^0] \quad \text{(Assumption 1)}$$

$$= E[Y_{j,1,2023}^1 - Y_{j,1,2022}^0] - E[Y_{j,0,2023}^0 - Y_{j,0,2022}^0] \quad \text{(Assumption 2)}$$

$$= E[Y_{j,1,2023}^1 - Y_{j,1,2022}^0] - E[Y_{j,1,2023}^0 - Y_{j,1,2022}^0]$$

$$+ E[Y_{j,1,2023}^0 - Y_{j,1,2022}^0] - E[Y_{j,0,2023}^0 - Y_{j,0,2022}^0]$$

$$= E[Y_{j,1,2023}^1 - Y_{j,1,2022}^0] - E[Y_{j,1,2023}^0 - Y_{j,1,2022}^0]$$

$$+ E[Y_{j,0,2023}^0 - Y_{j,0,2022}^0] - E[Y_{j,0,2023}^0 - Y_{j,0,2022}^0] \quad \text{(Assumption 3)}$$

$$= E[Y_{j,1,2023}^1 - Y_{j,1,2022}^0 - Y_{j,1,2023}^0 + Y_{j,1,2022}^0] = E[Y_{j,1,2023}^1 - Y_{j,1,2023}^0] = \theta_{j,2023}$$

Similarly, $\hat{\theta}_{j,2024}$ is unbiased for $\theta_{j,2024} - \theta_{j,2023}$ for all $j$:

$$E[\hat{\theta}_{j,2024}] = E[(Y_{j,1,2024} - Y_{j,1,2023}) - (Y_{j,0,2024} - Y_{j,0,2023})]$$

$$= E[Y_{j,1,2024}^1 - Y_{j,1,2023}^1] - E[Y_{j,0,2024}^0 - Y_{j,0,2023}^0] \quad \text{(Assumptions 1/2)}$$

$$= E[Y_{j,1,2024}^1 - Y_{j,1,2023}^1] - E[Y_{j,1,2024}^0 - Y_{j,1,2023}^0] \quad \text{(Assumption 3)}$$

$$= E[Y_{j,1,2024}^1 - Y_{j,1,2024}^0] - E[Y_{j,1,2023}^1 - Y_{j,1,2023}^0] = \theta_{j,2024} - \theta_{j,2023}$$

*Placebo Test and Bias Analysis*

If Assumption (3) is replaced with the equivalent parallel trends assumptions for $t < 2023$, then under the three assumptions, $\hat{\theta}_{j,t}$ would be an unbiased estimator of 0 for any $t < 2023$. So we calculate these estimators to use as placebo tests of this assumption. If the pre-2023 estimates are scattered around 0 and relatively small in magnitude, that suggests that Assumption 3 is reasonable. If a longer time series were available, a p-value could be calculated using these as a null distribution; see, e.g., Kennedy-Shaffer (2022).

In this case, the 2021 and 2022 estimates tend to be somewhat negative. This may indicate a negative pre-trend. If this were to hold true in 2023 in the absence of the intervention, that would indicate the obtained estimate is biased downward (toward the null).

*Interpretation of Estimand*

The estimands targeted, $\theta_{j,1,2023}$ and $\theta_{j,2024} - \theta_{j,2023}$, represent the average effect of the infield shift ban in 2023 and 2024 on the outcome $j$ for left-handed batter PAs with no runners on, compared to if no ban had been implemented. To estimate the effect on overall league statistics, then, assuming the ban affects no other PAs, we multiply the estimand by the percentage of PAs in this category. See Markes et al. (2024) for a similar approach to re-scaling estimates in this setting.

### *Analysis 2: Player-Specific Effects (Synthetic Control Method)*

*Notation and Covariates*

Again, let $j$ denote the outcome of interest (i.e., OBP, OPS, or wOBA) and $t$ the season from 2015 to 2024, excluding 2020. Index by $n = 1, \dots, 30$ the high-shift (target) players and by $m = 1, \dots, 58$ the low-shift (control/donor) players. Let $Y_{j,1,n,t}$ be the outcome for target player $n$ and $Y_{j,0,m,t}$ the outcome for control player $m$. Let the potential outcomes be defined analogously to Analysis 1 as $Y_{j,1,n,t}^1$ and $Y_{j,1,n,t}^0$ for the potential outcome for outcome $j$ of target player $n$ in season $t$ in the presence and absence, respectively, of the infield shift ban. We are interested in the effects of the ban for that player and outcome, $\theta_{j,n,t} = Y^1_{j,1,n,t} - Y^0_{j,1,n,t}$, for $t \geq 2023$.

For each player, we also record a set of covariates labelled $X_{k,1,n}$ and $X_{k,0,m}$ for target and control players, respectively, with $k = 1, \dots, K$ indexing the covariates. For the main analysis, the covariates are:

(1) Player age for 2022 season;

(2) Outcome $j$ value;

(3) Plate appearances;

(4) Hits;

(5) Singles;

(6) Home runs;

(7) Base-on-balls percentage;

(8) Strikeout percentage.

Item (2), the outcome value, is included for each pre-intervention season. For each of items (3)–(8), the values are included for the 2022 season, 2021 season, and the average of all pre-2020 seasons included in the analysis (seasons are included for which the target player has at least 250 PAs). Note that this means the seasons used in the covariates for the donor players depend on the target player for whom the analysis is conducted and the set of covariates depends on the outcome being analyzed since pre-intervention outcome values are included as covariates.

*Assumptions*

Assumptions (1) and (2) from the DID model in Analysis 1—consistency/no anticipation and no spillover—are required for SCM as well. The parallel trends assumptions is not required. Rather, the method relies on the target unit's pre-intervention characteristics (covariates $X_{k,1,n}$) being in the convex hull of those of the donor units so that the weight conditions can be approximately met. It also requires these weights to be stable over time (i.e., no systematic change in the ability of the donor units to reproduce the trajectory of the target unit) and that the effect be larger in magnitude than the volatility of the outcome. See Abadie (2021) for a more detailed discussion.

*Estimation*

The estimation procedure is described briefly here. For more details, see (Abadie 2021; Abadie et al. 2010, 2015). The implementation uses the tidysynth R package (Dunford 2023) and can be found at the GitHub repository ([https://bit.ly/QE-Baseball](https://bit.ly/QE-Baseball)).

For the main analysis, the estimate for outcome $j$ for target player $n$ in 2023 is given by:

$$\hat{\theta}_{j,n,2023} = Y_{j,1,n,2023} - \hat{Y}^0_{j,1,n,2023},$$

where

$$\hat{Y}^0_{j,1,n,2023} = \sum_{m=1}^{58} w_{j,n,m}\, Y_{j,0,m,2023}$$

is the synthetic control estimate for the potential outcome if the rule had not been implemented.

*Computing Weights*

The weights for outcome $j$, target player $n$, $\boldsymbol{w}_{j,n} = \left(w_{j,n,1}, \dots, w_{j,n,58}\right)$, are estimated by the SCM with the constraint that each weight must be nonnegative and the weights sum to 1: $\sum_{m=1}^{58} w_{j,n,m} = 1$. Note that a different set of weights is estimated for each outcome $j$ and for each target player $n$. The size of the donor pool also changes for each target player, since only donor players who had at least 250 PAs in the seasons for which that is true for the target player are included. This can be thought of as forcing the weight to be 0 for any donor player who does not meet this condition for a particular target player.

Within these constraints, the weights are chosen to minimize the root mean squared error between the weighted donor covariates and the target covariates. I.e., weights are chosen to minimize:

$$\left( \sum_{k=1}^{K} v_k \left[ X_{k,1,n} - \sum_{m=1}^{58} w_{j,n,m} X_{k,0,m} \right] \right)^{1/2},$$

where $v_1, \dots, v_K$ are importance weights on the covariates. These importance weights, in turn, are chosen by validation. They are the set of (nonzero, summing to 1) weights that minimize the mean squared prediction error of the corresponding SC results in some set of the pre-intervention periods. That is, if we write $w_{j,n,m}(\boldsymbol{v})$ as the SC weight for outcome $j$, target player $n$, donor player $m$, if the vector $\boldsymbol{v}$ of importance weights is used for the covariates, we select the $\boldsymbol{v}$ that minimizes for some set of pre-intervention periods $\tau$:

$$\sum_{t \in \tau} \left( Y_{j,1,n,t} - \sum_{m=1}^{58} w_{j,n,m}(\boldsymbol{v}) Y_{j,0,m,t} \right)^{1/2}.$$

Again, this is done for each target player and outcome, so the importance weights can vary across $j, n$ as well. Given the low number of pre-intervention periods available here, we use for $\tau$ the full set of available pre-intervention seasons (2015–2022, excluding 2020) where the target player had at least 250 PAs.

Alternate specifications could validate on a smaller set of time periods and/or exclude individual seasons' values of the outcome. This might be appropriate if data with less variability (potentially including bat-tracking data) were used, or a longer time series were available, as for specific individual players.

*Extension to 2024 Data*

To estimate the effect in 2024 as well, the analysis can be conducted similarly to the process described above. However, we modify the target player list and donor pool to exclude players with fewer than 250 PAs in 2024 to maintain comparability. This analysis yields estimates $\hat{\theta}_{j,n,2023}$ and $\hat{\theta}_{j,n,2024}$ for each outcome $j$ for each of the 27 target players $n$ who had at least 250 PAs in each of 2023 and 2024. Note that because of the slightly different donor pool, the estimate $\hat{\theta}_{j,n,2023}$ using this method may not exactly match that in the main analysis.

This could also be done to estimate $\hat{\theta}_{j,n,2024}$ alone by including all target and donor players with at least 250 PAs in 2024 regardless of their number of PAs in 2023. This is done as an additional sensitivity analysis and the results included in the GitHub repository. Only one additional target player is included with this weaker restriction.

*Placebo Tests*

To estimate the null distribution of estimates, we compute the SCM estimator for each of the 58 donor players, using the same procedure as above but leaving them out of the donor pool. For hypothesis testing for a single SCM estimate, the estimate, ratio of post-intervention to pre-intervention MSPE, or another test statistic for the target unit's estimate can be compared to the distribution of that test statistic from the placebo estimates (Abadie et al. 2010). However, with multiple target players here, we do not recommend using this as a strict hypothesis test. Rather, it can give a sense of the null distribution of estimates.

*Sensitivity Analyses*

Similarly, sensitivity analyses are conducted where we expect there to be no or minimal effect.

An in-unit sensitivity analysis is conducted using 2023 as the outcome, the same set of donor players, and the 25 players with 2022 shift rates between 15 and 30% as the "target" players. These players would be expected to experience a noticeably smaller effect of the ban than the high-shift players in the main analysis. The estimation procedure is the same as that described above, using this new set of "target" players.

An in-time sensitivity analysis is conducted, using the same analysis as if 2022 was the intervention year. The same donor and target players as the main analysis are used. The procedure is altered only by removing 2022 as a pre-intervention year and removing all covariates that rely on 2022 data. Player age in the 2021 season is used instead, just for maximum comparability of method.

*Interpretation of Estimands*

The estimates $\hat{\theta}_{j,n,2023}$ and $\hat{\theta}_{j,n,2024}$ target ATT estimands for outcome $j$ for the specific target player $n$ in the year 2023 and 2024, respectively. That is, they estimate the effect of the infield shift ban on the specific player in that specific year on that specific outcome. Averages could be constructed across a set of the target players, but they may not have a natural interpretation. Extrapolation to other similar players can be made with caution.